# DEPRENYL, AN OLD DRUG WITH NEW ANTICANCER POTENTIAL: MINI REVIEW

Nelson Durán[1,2], João C.C. Alonso[1,3] and Wagner J. Fávaro[1]

[1]Institute of Biology, Laboratory of Urogenital Carcinogenesis and Immunotherapy, Department of Structural and Functional Biology, University of Campinas, Campinas, SP, Brazil
[2]Nanomedicine Research Unit (Nanomed), Federal University of ABC (UFABC), Santo André, SP, Brazil.
[3]Paulínia Municipal Hospital, Paulínia, São Paulo, Brazil.

**ABSTRACT.**
The anticancer potential of monoamine oxidase (MAO) was observed in pre-clinical assays conducted with cell cultures and animals. L-Deprenyl (DEP) causes apoptosis in melanoma, leukemia and mammary cells. High-dose DEP has shown toxicity in mammary and pituitary cancers, as well as in monoblastic leukemia, in rats. DEP accounts for immune-stimulant effect capable of increasing natural killer cell activity, IL-2 generation, as well as of inhibiting tumor growth. DEP administration in old female rats has increased IL-2 generation and inverted the age-related depletion of IFN-γ generation in the spleen. Co-adjuvant DEP administration helped preventing/mitigating symptoms associated with peripheral neuropathy in cancer treatment. It also enhanced the cytotoxic effects of antineoplastic drugs - such as doxorubicin, cisplatin, among others - in cancer cells while they protected healthy cells from being damaged. DEP presented effect against dysfunctions such as debilitating hormone imbalance triggered by pituitary gland tumor; this gland produces the stimulatory hormone of adrenocorticotropic hormone which was related to the exacerbation of this disease. Thus, DEP emerges as an excellent potential drug against several cancer types and it also presents low toxicity in Parkinson`s disease patients subjected to long treatment with it.

# 1. Introduction

(R)-N-methyl-N-(1-phenylpropan-2-yl)-prop-1-yn-3-amine or (*R*)-(−)-N,α-dimethyl-N-(2-propynyl) phenethylamine [(−)-deprenyl or L-deprenyl (DEP)] (Fig.1) have proven to be an effective selective blocker of monoamino oxidase B (MAO-B) activity in the brain of mammals. They prolong the lives of rats and of many other animals, such as dogs and hamsters, as well as acts in *Drosophila melanogaster*.[1]
Selegiline, Eldepryl, Jumex, Zelapar, Carbex, Atapryl and Deprenalin are commercial names attributed to DEP, which is mainly used against Parkinson's disease.
DEP was qualified as the first specific B-type monoamine oxidase (MAO-B) inhibitor, which was defined as catecholaminergic-activity enricher or enhancer (CAE) drug. It is necessary keeping in mind that tryptamine is primarily a natural serotonergic neuron enricher that enables (2R)-1-(1-benzofuran-2-yl)-N-propylpentane-2-amine (BPAP) (Benzofuranyl propylamino pentane) invention (Fig.1). This substance, in its turn, is a powerful synthetic product used for this biological activity type.[2] Extensive reviews about the nature of DEP as enhancer regulator, as well as and strategies focused on synthetic-enhancer substance progress have been published.[1, 3-6]

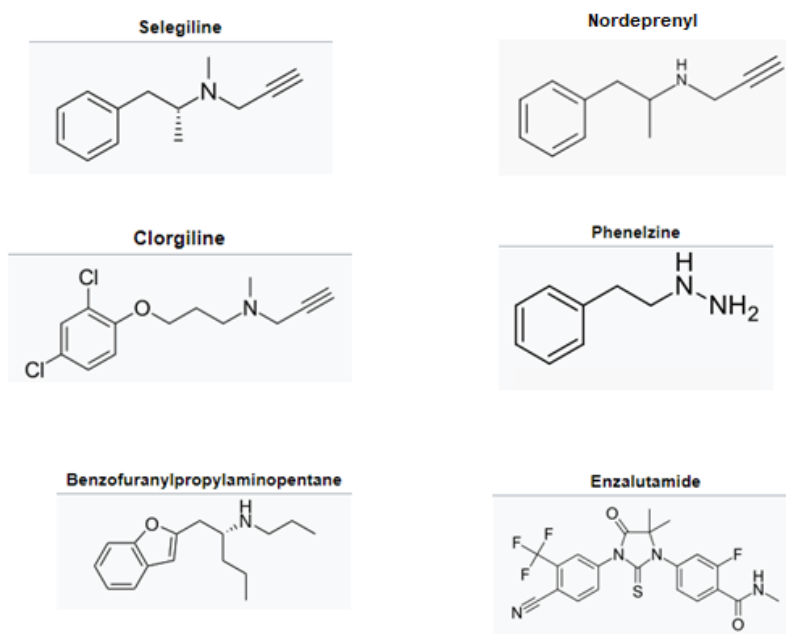

**Figure 1. Chemical structures of compounds in the text.**

## 2. Cancer perspectives

### *2.1. General aspects*

The feasible anticancer activity of MAO inhibitors was already indicated in several reports. High MAO content was observed in several cancer cells and the repression of this oxidase resulted in antiproliferative action.[7,8] Almost all the aforementioned results were observed in assays conducted *in vitro*, whereas few of them were observed in studies conducted with animals. However, it is essential emphasizing that phenelzine, which is a nonselective MAO inhibitor, is nowadays at phase-II clinical trial for non-metastatic reiterative prostate cancer therapy [ClinicalTrials.gov Identifier: NCT02217709-USA, in progress], a fact that suggests the clinical importance of MAO inhibitors associated with anticancer activity.

### *2.2. Cell cultures*

High-dose DEP has shown cytotoxic effects on several cancer cell lines *in vitro*. DEP application at concentration of $10^{-3}$ M was capable of inducing apoptosis in melanoma cell line.[9,10] This report has investigated the cytoprotective action of MAO-B inhibitor DEP in A-2058 human melanoma cells cultured in RPMI/fetal calf serum. Serum deprivation after 24 h has triggered apoptosis in cultured cells that may have decreased in number due to $10^{-9}$–$10^{-13}$ M DEP application. DEP metabolites, such as desmethyl-DEP (Nordeprenyl) (Fig.1), were not successful in triggering the same action. The anti-apoptotic activity of DEP was avoided by concomitant administration of the microsomal drug-metabolizing enzyme inhibitor known as SKF-525A. Data have shown that DEP requires metabolic change to become anti-apoptotic. On the other hand, low ($10^{-13}$ M) DEP and Nordeprenyl doses have caused apoptosis in non-serum-deprived A2058 cell culture. The SKF-525A inhibitor did not avoid the apoptosis-inducing action of DEP, and this outcome suggests that metabolic transformation is not necessary to enable such an action. High ($10^{-3}$ M) DEP dose has led to extremely high Caspase 3 activity in non-serum-deprived A-2058 cell culture, however, low ($10^{-9}$M to $10^{-13}$ M) DEP doses have kept Caspase 3 activity in its original status in case of serum-deprivation.[10]

Acute myelogenous leukemia (AML) is defined as hematopoietic cancer presenting high mortality rates - more than 50% of patients with this disease do not react to chemotherapy.[11] Although therapeutic formalities have evolved enough to help managing AML, the variety of gene mutations and the heterogeneity in some cancer cells have significantly impaired efforts focused on enabling their full remission.[12,13] The object of study in AML research has progressively changed from gene requirement to organelle-based therapy in order to overcome disagreements in recognizing operative drug targets.[14]

MAO-A and MAO-B activity in AML cells was low (BMMNCs and KG-1α cells.), and these results have indicated that the anticancer action of DEP did not depend on MAO-B activity. According to the aforementioned outcomes, DEP administration ($10^{-3}$M) has diminished the mRNA level of mitochondrial respiration and glycolysis-related genes, which is a different activity from that of the MAO-B inhibitor.[15]

The effect of MAO-B action in bone marrow cells and in the whole brain tissue was analyzed, since DEP can selectively inhibit MAO-B. The whole brain presented MAO-B activity, whereas bone marrow cells did not. Previous studies had mentioned that cancer cells (glioma) showed increased MAO-B levels.[16]

Leukemia KG-1α and HL-60 cell lines were tested in order to check whether bone marrow-derived cancer cells present MAO action. Data did not show difference between the presence or absence of MAO-A or MAO-B inhibitors, and it indicated that leukemic cell lines have accounted for these activities. Such data have indicated that the toxic activity of DEP in KG-1α cells did not depend on oxidase activity inhibition.[15]

Reactive oxygen species (ROS) had toxic effect on cancer cells (e.g. mammary tumor) subjected to high-dose DEP treatment.[17,18] Mitochondrial oxygen consumption rate (OCR) was inhibited by DEP treatment applied to KG-1α cells for 24h, because mitochondrial oxidative phosphorylation complexes I and III are the main ROS production sites. Although mitochondrial ATP generation decrease was rapidly compensated by glycolytic pathway upregulation to maintain the intracellular ATP rate, the extracellular acidification rate (ECAR), which leads to intracellular lactate production, was not affected by it.[19]

Lajkó et al.[17] have demonstrated that besides its cytotoxic effect, DEP can have the potency to pull raised adhesion and has chemorepellent action in monocyte models (e.g., Mono Mac 6 cell line deriving from monoblastic leukemia). Moreover, the inductive adhesion, chemorepellent action and useful cytotoxic activity of DEP, and of some other similar products, point towards their likely inhibitory action in metastasis production in primary tumors.

### 2.3. Pre-clinical assays with animals

High-dose DEP therapy has toxicity in rats' monoblastic leukemia, pituitary and mammary cancer cells, as observed by monitoring the incidence of serum prolactin in these cancer types.[17, 20-22] Besides its activity in solid tumors, DEP was capable of reducing the incidence of monoblastic leukemia cells by inducing interferon-γ and norepinephrine generation, as well as CD8+ T lymphocytes and NK cell clusters in rats' spleen.[20,22-24]

As previously mentioned, high DEP concentrations had anticancer effects on rats, since they have effectively reduced both breast tumor size and tumor cell viability. There is evidence that DEP was capable of diminishing the action of an estrogen-receptor (ER)-dependent intracellular signaling route in ER-positive human breast cancer.[25]

First evidences have shown DEP as the first β-phenylethylamine (PEA)-derived synthetic catecholaminergic activity enricher (CAE) product, although it presents different markers from those of uncommon enhancer-sensitive rat brain regulators.[1] Results have shown that 20/40 of saline-treated rats presented fibromyxosarcoma. Groups of rats treated with 1 µg/kg DEP presented 15/40 individuals with this very same sarcoma, whereas groups treated with 100 µg/kg DEP presented 11/40 individuals with it. However, a tryptamine-derived (2R)-1-(1-benzofuran-2-yl)-N-propylpentane-2-amine (Benzofuranylpropylaminopentane-BPAP) (Fig.1) enabled better results in 7/40 rats at the dose of 50 µg/kg. Connection to different places in vesicular monoamine-transporter-2 (VMAT2) was the main inducing product-activation mechanism capable of elucidating the extensive bi-modal property and bell-shaped concentration effect of DEP curves. The rescue of inductor products and the evidence that DEP (distinct markers of unrecognized brain enhancer sensitive regulations) showed the action of enhancer-sensitive tumor-manifestation-suppressing (TMS) regulation in rat brain. Knoll et al.[1] reported to the need of testing a low DEP dose in humans to avoid malignant tumor dissemination. Other studies have shown that DEP protected neurons from several neurotoxic agents and increased neurotrophins production (brain-derived neurotrophic factor, nerve growth factor and glial cell-derived neurotrophic factor), which are natural neuron-defensive agents. In addition, it presented immune-stimulant effect, increased IL-2 and natural killer cell activity, suppressed tumor growth, serum prolactin and brain monoamine metabolism in rats with carcinogen-induced mammary tumors, and was associated with induced immune action and with central and peripheral neurotransmission processes.[5]

Decreased IL-2 generation by splenocytes was not observed in female rats younger than 21 months, but it was observed in female rats aged up to 24 months.[18,25,26] This outcome was attributed to age difference amid female rats selected for these studies.[18,25,26] DEP treatment applied to old female rats has increased IL-2 generation and reversed age-related impairment in IFN-γ output in the spleen.[27]

Similar DEP immune-stimulatory properties were observed in the spleens of old male F344 rats and tumor-bearing rats, as well as in peripheral blood mononuclear cells of healthy humans, who presented increased pro-inflammatory cytokines, IL-1 and IL-6 production.[28,29]

## 3. Co-adjuvant effect

An invention[30] based on DEP to help preventing or mitigating symptoms associated with peripheral neuropathy in cancer treatment was patented. This invention is a method focused on protecting patients with peripheral neuropathy, caused by toxic agents, through the administration of a proper DEP dose capable of diminishing or ruling out one, or more, symptoms associated with such a neuropathy. DEP was effective against any toxic chemotherapeutic agent capable of producing peripheral nerve dysfunction. It was mostly effective against agents that trigger acute neuropathic side effects, such as cisplatin, paclitaxel and vincristine. A patient with endometrial carcinoma, was subjected to intravenous bolus injection of Vincristine on a weekly basis. The toxic action of vincristine has produced sensory loss in patient's fingers/toes, ankle jerk reflex and weakness/ postural hypotension. The patient was given DEP (orally) twice a day and all toxic effects have reduced. A patient with ovarian cancer was subjected to Cisplatin injections on a weekly basis and to oral DEP doses twice a day. The effect of the therapy on tumor progression was evaluated and treatment effectiveness was discussed. A patient with breast cancer was subjected to oral DEP doses for several cycles as long as the useful effect on tumor advances was obtained or inadmissible side effect was not eliminated

Antineoplastic drug modulators, such as DEP, have enhanced the cytotoxic effects of antineoplastic drugs (e.g., doxorubicin, cisplatin, among others) on cancer cells and protected healthy cells from being damaged. Modulators can be used to enhance the selectivity and availability of conventional antineoplastic drugs, to diminish unwanted cancer chemotherapy side-effects, to improve cancer chemotherapy effect, to enhance the treatment of cancer patients for whom other procedures are ineffective, to improve the therapy of cancer patients who are non-responsive or poorly responsive to medicine-resistance and/or toxicity restricted treatment regimens and to provide efficient chemotherapy for, so far, untreatable cancers.[31]

Cushing's disease is a subtle hormone instability that takes place when the adrenal glands produce excessive cortisone, 85% of such production is caused by tumor in the pituitary gland, which segregates a stimulatory hormone known as adrenocorticotropic hormone (ACTH). Approximately 80% of dogs with cognitive dysfunction present health condition improvements after one month of DEP administration therapy.[32]

## 4. Mechanistic aspects of DEP in immunology and cancer

MAO catalyzes the oxidative deamination of monoamine neurotransmitters and dietary amines. MAO-A and MAO-B, also called isoenzymes, are two enzymes presenting different substrate and inhibition features, but they have equal intron-exon organization. Interestingly, these isoenzymes present different behavior in mice. MAO-A knockout (KO) mice display aggressive behavior, the first clear proof of connecting genes to conducts. Unlike MAO-A, MAO-B KO were non-aggressive in mice and averse to Parkinsongenic neurotoxin. Later on, it was determined that MAO-A was excessively expressed in prostate cancer and that such a behavior followed the malignancy stage. Oncogenic mechanistic aspects involve a ROS-activated AKT/FOXO1/TWIST1 signaling pathway. MAO-A KO in which MAO-A was removed has reduced prostate cancer stem cells and ruled out invasive adenocarcinoma and improved standard chemotherapy.[7] A near-infrared (NIR) dye-conjugated clorgyline (MAO-A inhibitor) (Fig.1) was produced in this field to be used as new theranostic (therapeutic/diagnostic) agent against cancer. However, MAO-B was excessively expressed in glioma and non-small cell lung cancer. Researchers have suggested that MAO-A and MAO-B function in many cancer types enables cancer treatment to take new directions.[7] Remarkably, MAO-A and MAO-B activities were low in acute myelogenous leukemia (AML) cells and such findings have indicated that the anticancer action of DEP in AML did not depend on MAO-B.[15]

Gaur et al.[8] observed that MAO-A inhibitors, mainly clorgyline and phenelzine (Fig.1), were effective in diminishing MAO-A activity in human prostate cancer cells. These inhibitors have effectively diminished LNCaP, C4-2B, and 22Rv1 cell growth and induced additive growth-inhibitory action when they were associated with enzalutamide (Fig.1). Clorgyline has suppressed AR-FL and AR-V7 expression in 22Rv1 cells and has effectively diminished enzalutamide-resistant C4-2B cell line growth and led to increased AR-V7 expression. Thus, these inhibitors can also diminish the growth and proliferation of androgen-sensitive and castration-resistant prostate cancer cells. These outcomes provide preclinical validation to inhibitors, either alone or in association with antiandrogens, to be used for therapeutic purposes in patients with acute prostate cancer.

Based on previous reports, the microsomal drug-metabolizing enzyme inhibitor known as SKF-525A, in association with low DEP concentrations, has prevented the anti-apoptotic action of this compound and showed that metabolic DEP conversion was necessary to achieve anti-apoptotic action. On the other hand, treatment of non-serum-deprived A-2058 cell cultures with high ($10^{-3}$ M) DEP dose and its known metabolites resulted in significant increase in apoptotic cell death rate. The SKF-525A inhibitor did not potentiate or prevent apoptosis caused by DEP at such a high concentration. Data have strongly indicated that metabolism is not necessary to enable pro-apoptotic action at high DEP dose, as well as that Caspase 3 was associated with apoptosis (activated by high DEP dose) and that Caspase 3 activity did not reach levels higher than the non-serum-deprived control level, when DEP was applied to serum-deprived A-2058 cultures, at very low doses. Results have indicated that unknown DEP metabolites show anti-apoptotic effect by inhibiting Caspase 3. Data have shown that both the anti-apoptotic and apoptosis-inducing actions of DEP were very concentration-dependent. The aforementioned factor could be important and have effect on cancer treatment, since it enables neuronal-origin tumor therapies based on high DEP doses.[10] In addition, DEP concentrations of $10^{-9}$ to $10^{-13}$ M were capable of suppressing neurotoxin-induced apoptosis, although DEP concentration of $10^{-3}$ M has caused apoptosis in neuro-ectodermal source tissues.[33]

DEP administration in isolated BMMNCs (AML cells) of FLT3-ITD knock-in mice has shown diminished cell viability in a dose-dependent way (0.5-4.0 x $10^{-3}$M). Measurements of DEP power over immature leukocyte cells and, more specifically, over isolated BMMNC cell viability in FLT3-ITD knock-in and WT mice, have shown that DEP concentration of 2 mM had stronger effect (~two-fold) on isolated BMMNCs in FLT3-ITD knock-in mice than that in WT mice. Cell viability was evaluated in KG-1α cells (AML cell lines), to feature cell-death type induced by DEP in leukemia cells. Similarly to the outcomes recorded for BMMNCs, KG-1 α cell viability has decreased in a dose-dependent way. Moreover, increased breakage of poly [ADP-ribose] polymerase 1 (PARP-1) and caspase-3 (apoptosis markers) was observed at 4.0 x $10^{-3}$ M DEP. These outcomes have indicated that late treatment with DEP led to apoptotic cell death in BMMNCs and KG-1α cells.[15,19]

Based on these findings, it was possible seeing that DEP produced apoptotic cell death in AML synchronous due to mitochondrial OCR and cytosolic ECAR suppression, a process that is not influenced by MAO-B suppression *in vitro* and *ex vivo*. DEP is a likely nominee for first-line AML treatment in the near future, when it comes to the relevance of drug development progress to help managing intracellular metabolic flux.[15]

Mammary tumor-bearing rats have shown lower splenic IL-2 and IFN-γ levels, as well as lower splenic norepinephrine (NE) concentration and hypothalamic dopaminergic activity than healthy rats. On the other hand, the administration of 2.5-5.0 mg/kg of DEP has decreased the size and number of mammary tumors. DEP has led to tumor regression, increased immune responses in the spleen, as well as improved IL-2 and IFN-γ generation, and NK cell action. Neural parameters improved by DEP administration comprised splenic norepinephrine (NE) concentration in the spleen and tuberoinfundibular dopaminergic (TIDA) neuronal action in the hypothalamus. Outcomes have indicated that mammary tumorigenesis was associated with sympathetic noradrenergic (NA) action suppression in the spleen, TIDA action in the hypothalamus and with cell-mediated immunity, whereas inverted suppression of catecholaminergic neuronal actions in the central and peripheral nervous systems by DEP has likely induced anti-tumor immunity.[24]

Clinical studies available in the literature have reported that DEP assumingly diminishes neurological deficiency progression in Parkinson's disease patients, as well as cognitive decay in Alzheimer's disease ones. With respect to animal trials, DEP was capable of increasing the survival rate of immunosuppressed mice. The immune response to any infection or chronic inflammatory process is guided by the enhanced synthesis of cytokines such as IL-1ß and subsequent IL-6. These brain infection types are strongly correlated to levels of cytotoxic cytokines such as tumor-necrosis-factor-alpha (TNF).[28] DEP control in IL-1ß, IL-6 and TNF biosynthesis in peripheral blood mononuclear cells (PBMC) from healthy human blood donors was investigated in PBMC cultured at different DEP concentrations. DEP has increased IL-1ß and IL-6 synthesis and reduced TNF production. Data has indicated that DEP control on cytokine biosynthesis likely add to its putative neuroprotective qualifications.[28]
ThyagaRajan et al.[34] had suggested likely correlation between some immune-senescence parameters and the age-related turn-down in sympathetic noradrenergic nerve fibers in the spleen and lymph nodes of F344 rats. The action of natural killer cells and Con A-induced IL-2 generation have increased in old DEP-treated rats in comparison to the untreated ones - no changes were observed in CD5+ T-cell contents. Besides the transformation in immune responses, norepinephrine rate and the volume density of noradrenergic and neuropeptide-Y nerve fibers have partially increased in the spleen of old DEP-treated rats. Unlike the treatment conducted *in vivo*, DEP addition *in vitro* did not lead to any change in the Con A-induced IL-2 segregated by splenocytes in old rats. Thus, data have indicated that DEP's ability to improve certain immune responses was interlinked to the repair of sympathetic noradrenergic and neuropeptide-Y nerve fibers in the spleen of old rats.[29]

## 5. Final remarks

Exacerbated MAO expression in several cancer cells was observed and oxidase inhibition resulted in antiproliferative effect. However, studies conducted *in vitro* and *in vivo* have suggested that the anticancer effect of DEP, as MAO-B inhibitor, was not dependent on MAO-B itself.
Prolonged treatment with DEP has induced apoptotic cell death in acute myelogenous leukemia (AML) cells and increased cleaved PARP-1 and caspase-3 (apoptosis markers). DEP-induced mammary tumor regression in rats presented increased immune parameters in the spleen, if one takes into consideration the enhanced IL-2 and IFN-γ generation and NK cell action.
Besides its anticancer activity, DEP plays an important role as co-adjuvant drug in cancer treatment. It was capable of mitigating symptoms associated with peripheral neuropathy in cancer patients and was effective against agents with particularly severe neuropathic side effects such as cisplatin, paclitaxel and vincristine. Antineoplastic drug modulators, such as DEP, were capable of enhancing the cytotoxic effects of antineoplastic drugs (e.g., doxorubicin, cisplatin, among others) on cancer cells and of protecting healthy cells from being damaged.
Finally, all these studies conducted with animals, both *in vitro* and *in vivo,* have shown the great potential of DEP to be used for cancer treatment purposes. DEP is safe and has been used for many years in Parkinson´s disease patients, since it presents very low toxicity in humans, which assures its effectiveness to be used for this new purpose (cancer treatment).

**ACKNOWLEDGEMENT**

The authors would like to thank the São Paulo Research Council (FAPESP grant 2018/10052-1), the Brazilian National Council for Scientific and Technological Development (CNPq grant 552120/2011-1) and Coordinating Agency for Advanced Training of Graduate Personnel (CAPES – funding code:
001).